\documentstyle[11pt, amssymb]{article}

\begin{document}

\begin{flushleft}
{\Large {\bf Families of KP solutions associated with tropical curves having nontrivial weights}}
\end{flushleft}

\begin{flushleft}
{\large {\bf Takashi Ichikawa}} 
\end{flushleft}

\begin{flushleft} 
Department of Mathematics, Faculty of Science and Engineering, 
Saga University, Saga 840-8502, Japan. E-mail:  ichikawn@cc.saga-u.ac.jp 
\end{flushleft} 



\noindent
{\bf Abstract:} 
In order to generalize a program by Agostini and others producing new KP solutions, 
we construct families of quasi-periodic KP solutions which are derived from 
degenerating Riemann surfaces associated with tropical curves having nontrivial weights. 
By taking the regularized tropical limits of these solutions, 
we obtain formulas of general KP solutions 
which are expressed by mixtures of solitons and Riemann theta functions. 



\section{Introduction} 

Recently, tropical geometry was applied to studying soliton solutions to 
the KP hierarchy in the works of Kodama-Williams \cite{KoW}, 
Kodama \cite{Ko1}, Agostini-Fevola-Mandelshtam-Sturmfels \cite{AgFMS} 
and others. 
Especially, the behavior of line solitons was analyzed in \cite{KoW, Ko1} 
by taking their tropical limits, 
and a program given in \cite{AgFMS} will construct solitons as tropical limits of 
real theta functions associated with periods of tropical curves having trivial weights. 
In \cite{I2}, this program was realized by constructing families of 
quasi-periodic {\it KP solutions,} 
namely solutions to the KP hierarchy associated with these tropical curves, 
and these regularized tropical limits are KP solitons  
which were shown to contain all line solitons 
by Nakayashiki \cite{N1, N2}, Abenda-Grinevich \cite{AG} and Kodama \cite{Ko2}. 

The aim of the present paper is to extend results of \cite{I1, I2} 
to general tropical curves, namely tropical curves with nontrivial weights. 
First, we show a variational formula for the periods of degenerating Riemann surfaces 
associated with general tropical curves in which and results of \cite[3.3]{I2} 
the divergent term and constant term are explicitly given, 
while variational formulas of Fay \cite{F}, Bainbridge-M\"{o}ller \cite{BM} and 
Hu-Norton \cite{HuN} shown by analytic methods 
do not describe the constant terms explicitly. 
Second, by using our variational formula, 
we construct families of quasi-periodic KP solutions from general tropical curves 
which are shown to have regularized tropical limits as KP solutions expressed by 
mixtures of solitons and Riemann theta functions. 
This construction also suggests the existence of a compactified (or extended) Torelli map 
sending degenerate algebraic curves with tropical structure on their dual graphs to  compactified Jacobians which will extend results of Mumford and Namikawa 
\cite[Section 18]{Nam} treating only tropical structure with constant length functions. 
Similar KP solutions of mixed type were obtained by Nakayashiki and others \cite{BeEN, N3} 
using degenerations of restricted types for Riemann surfaces which are hyperelliptic or trigonal, 
and by Nakayashiki \cite{N4} as solitons on elliptic backgrounds 
using the action of vertex operators on the KP hierarchy 
(see also Kakei \cite{K} and Li-Zhang \cite{LZ}). 
Then it is interesting to clarify the relationship between these KP solutions and ours.

\section{Quasi-periodic KP solutions}

Fix a Riemann surface $R$ of genus $g > 0$ 
with symplectic basis $\{ a_{i}, b_{i} \}_{1 \leq i \leq g}$ of $H_{1}(R, {\mathbb Z})$, 
a point $p \in R$ and a local coordinate $z$ at $p$ such that $z(p) = 0$, 
and denote the whole data by $X = \left( R, \{ a_{i}, b_{i} \}, p, z \right)$ 
which is called a {\it dressed Riemann surface}. 
Then there exists a unique basis $\{ \omega_{1},..., \omega_{g} \}$ consisting of 
holomorphic abelian differentials, i.e., $1$-forms on $R$ 
which is normalized for $\{ a_{1},..., a_{g} \}$ 
in the sense that 
$$
\int_{a_{i}} \omega_{j} = \delta_{ij} \ (1 \leq i, j \leq g), 
\eqno(2.1) 
$$
where $\delta_{ij}$ denotes the Kronecker delta. 
Therefore, 
$$
Z(X) = \left( Z_{i,j}(X) \right)_{1 \leq i, j \leq g} = 
\left( \int_{b_{i}} \omega_{j} \right)_{1 \leq i, j \leq g} 
$$ 
is the period matrix of $\left( R, \{ a_{i}, b_{i} \} \right)$. 
Take $r_{j,m}(X) \in {\mathbb C}$ such that 
$$
\omega_{j} = \sum_{m=1}^{\infty} r_{j,m}(X) z^{m-1} dz \ \ \mbox{at $p$}, 
$$
and put $r_{m}(X) = \left( r_{1,m}(X),..., r_{g,m}(X) \right)$. 
By the Riemann-Roch theorem, for each positive integer $n$, 
there exists a unique meromorphic abelian differential $\omega^{(n)}$ on $R$ 
with unique pole of order $n+1$ at $p$ 
which is normalized for $(\{ a_{i} \}_{1 \leq i \leq g}, p, z)$ 
in the sense that 
\begin{itemize}

\item 
$\omega^{(n)}$ is holomorphic outside $p$, 

\item 
$\displaystyle \omega^{(n)} = 
\left( \frac{1}{z^{n+1}} + \sum_{m=1}^{\infty} \frac{q_{n,m}(X)}{n} z^{m-1} \right) dz$ 
at $p$ for some $q_{n,m}(X) \in {\mathbb C}$, 

\item 
$\displaystyle \int_{a_{i}} \omega^{(n)} = 0$ for any $i = 1,..., g$. 

\end{itemize}
We denote the Riemann theta function defined for $\left( R, \{ a_{i}, b_{i} \} \right)$ 
of $z = (z_{1},..., z_{g}) \in {\mathbb C}^{g}$ by 
$$
\Theta_{(R, \{ a_{i}, b_{i} \})} (z) = 
\sum_{u \in {\mathbb Z}^{g}} 
\exp \left( \pi \sqrt{-1} \left( u Z(X) u^{T} + 2 z u^{T} \right) \right) 
\eqno(2.2) 
$$
which is an absolutely convergent series since the symmetric matrix $Z(X)$ 
has positive definite imaginary part. 
Let $t_{m}$ $(m = 1, 2,...)$ be indeterminates, 
and put $t = (t_{1}, t_{2},...)$. 
Then following \cite[Definition 3.5]{KawNTY} (cf. \cite[(14)]{Alv-GGR}, \cite{IMO}, \cite[(24)]{V}),  
for the dressed Riemann surface $X$ and $c = (c_{1},..., c_{g}) \in {\mathbb C}^{g}$ 
corresponding to a divisor of degree $0$ on $R$ by the Abel-Jacobi map, 
the associated {\it tau function} $\tau(t, X_{c})$ is defined as 
$$
\tau (t, X_{c}) = \exp \left( \frac{1}{2} \sum_{n, m = 1}^{\infty} q_{n,m}(X) t_{n} t_{m} \right) 
\cdot \Theta_{(R, \{ a_{i}, b_{i} \})} \left( c + \sum_{m = 1}^{\infty} r_{m}(X) t_{m} \right). 
\eqno(2.3) 
$$
Then $\tau(t, X_{c})$ is regarded as an element of 
${\mathbb C}[[t]] = {\mathbb C} [[t_{1}, t_{2},...]]$ by the expression 
$$
\prod_{i = 1}^{g} \exp \left( 2 \pi \sqrt{-1} \sum_{m = 1}^{\infty} r_{i,m}(X) t_{m} \right)^{u_{i}}
= \sum_{n = 0}^{\infty} \frac{1}{n!} \left( 2 \pi \sqrt{-1} 
\sum_{m = 1}^{\infty} \left( \sum_{i = 1}^{g} u_{i} \cdot r_{i,m}(X) \right) t_{m} \right)^{n} 
$$
for $u_{i} \in {\mathbb Z}$. 
If $\Theta_{(R, \{ a_{i}, b_{i} \})}(c) \neq 0$, 
then we put
$$
\frac{\tau(t - [\alpha], X_{c})}{\tau(t, X_{c})} = 
1 + \sum_{k = 1}^{\infty} w_{k} (t, X_{c}) \alpha^{k}, 
\eqno(2.4) 
$$
where $[\alpha] = \left( \alpha, \alpha^{2}/2, \alpha^{3}/3,... \right)$, 
and define two micro-differential operators
$$
W (t, X_{c}) = 1 + \sum_{k = 1}^{\infty} w_{k} (t, X_{c}) \partial_{x}^{-k}
\eqno(2.5) 
$$
and
$$
L (t, X_{c}) = W (t + x, X_{c}) \cdot \partial_{x} \cdot W (t + x, X_{c})^{-1}
\eqno(2.6) 
$$ 
with coefficients in ${\mathbb C}[[x, t]]$, 
where $t + x = (t_{1} + x, t_{2}, t_{3},... )$. 
Then it is known (cf. \cite{Kr, SW}) that $L(t, X_{c})$ satisfies the KP hierarchy
$$
\frac{\partial L}{\partial t_{n}} = \left[ (L^{n})_{+}, L \right] \ \ (n = 1, 2,...). 
$$
In particular,
$$
u (x, t_{2}, t_{3})
= \frac{\partial^{2}}{\partial x^{2}} \log \Theta_{(R, \{ a_{i}, b_{i} \})} 
\left( c + x r_{1}(X) + t_{2} r_{2}(X) + t_{3} r_{3}(X) \right) +  q_{1,1}(X) 
$$
satisfies the KP equation
$$
\frac{3}{4} \frac{\partial^{2} u}{\partial t_{2}^{2}} - \frac{\partial}{\partial x} 
\left( \frac{\partial u}{\partial t_{3}} - \frac{1}{4} \frac{\partial^{3} u}{\partial x^{3}} - 
3 u \frac{\partial u}{\partial x} \right) = 0. 
$$ 
Since $L(t, X_{c})$ and $u(x, t_{2}, t_{3})$ are expressed by theta functions 
with quasi-periodicity, 
they are called {\it quasi-periodic KP solutions}.

\section{Variation of theta functions and KP solutions}

\subsection{Tropical curve and associated complex curves} 

Following \cite{M, MZ}, 
a tropical curve $C$ is defined as a connected graph $\Delta = (V, E)$ 
with length function $l : E \rightarrow {\mathbb R}_{> 0}$ and 
weight function $w : V \rightarrow {\mathbb R}_{\geq 0}$. 
In what follows, fix an orientation of each edge in $E$, 
and for an oriented edge $h \in \pm E$, 
denote by $v_{h} \in V$ its starting (initial) vertex and 
by $|h| \in E$ the corresponding nonoriented edge. 
For each $v \in V$, 
take a Riemann surface $R_{v}$ of genus $w(v)$ and marked points $x_{h}$ on $R_{v}$ 
with local coordinates $\xi_{h}$ at $x_{h}$ such that $\xi_{h}(x_{h}) = 0$, 
where $h \in \pm E$ satisfy $v_{h} = v$. 
Put $h_{1}(\Delta) = {\rm rank}_{\mathbb Z} H_{1}(\Delta, {\mathbb Z})$, 
and define a nodal complex curve $R_{0}$ of genus 
$$
g := h_{1}(\Delta) + \sum_{v \in V} w(v) 
$$
as the union of $R_{v}$ $(v \in V)$ 
by identifying $x_{e}$ and $x_{-e}$ for $e \in E$. 
Then we consider the deformation ${\mathcal R}_{C}$ of $R_{0}$ 
by sufficiently small complex parameters $y_{e} = y_{-e}$ $(e \in E)$ 
such that ${\mathcal R}_{C}$ is obtained from the relations $\xi_{e} \cdot \xi_{-e} = y_{e}$. 
For members of ${\mathcal R}_{C}$ associated with nonzero $y_{e}$ $(e \in E)$, 
take symplectic basis 
$$
\left\{ a_{v,i} \left( v \in V \right), a_{j}, \, b_{v,i} \left( v \in V \right), b_{j} \right\} 
\eqno(3.1)  
$$  
of their first integral homology groups which satisfies the following: 
\begin{itemize}

\item 
For each $v \in V$, 
$\left\{ a_{v,i}, \, b_{v,i} \right\}_{i}$ gives a symplectic basis of $H_{1}(R_{v}, {\mathbb Z})$. 

\item 
Each $a_{j}$ is represented as a small oriented closed path in the $\xi_{h}$-plane 
around $x_{h}$ for an oriented edge $h \in \pm E$, 
and $\{ b_{j} \}$ gives a basis of $H_{1}(\Delta, {\mathbb Z})$ 
when $y_{e} \rightarrow 0$ $(e \in E)$. 

\end{itemize}
Denote by $\omega_{C/S}$ the dualizing sheaf on a nodal curve $C$ over $S$ 
which is called the canonical invertible sheaf in \cite[Section 1]{DM}. 
The sheaf $\omega_{C/S}$ is invertible on $C$, and is functorial for $S$. 
Furthermore, 
if $S = {\rm Spec}(k)$ for an algebraically closed field $k$ and 
$\nu : C' \rightarrow C$ is the normalization of $C$ with points $p_{i}, q_{i} \in C'$ 
such that $\nu(p_{i}) = \nu(q_{i})$ are the double points on $C$, 
then $\omega_{C/S}$ is the sheaf of $1$-forms $\eta$ on $C'$ which are regular 
except for simple poles at $p_{i}, q_{i}$ such that 
${\rm Res}_{p_{i}}(\eta) + {\rm Res}_{q_{i}}(\eta) = 0$ (cf. \cite[Section 1]{DM}). 
We consider sections of the dualizing sheaf on ${\mathcal R}_{C}$ over 
$$
U_{r} = \left\{ \left. (y_{e})_{e \in E} \in {\mathbb C}^{E} \ \right| \, 
|y_{e}| < r \ (e \in E) \, \right\} 
$$ 
for a sufficiently small positive number $r$, 
and call global sections of $\omega_{{\mathcal R}_{C}/U_{r}}$ 
{\it stable abelian differentials} on ${\mathcal R}_{C}/U_{r}$. 
\vspace{2ex}

\noindent 
{\bf Theorem 3.1.} 
\begin{it}

{\rm (1)} 
There exists a unique basis $\{ \omega_{v,i} \left( v \in V \right), \omega_{j} \}$ 
of stable abelian differentials on ${\mathcal R}_{C}/U_{r}$ 
which is normalized for $\left\{ a_{v,i} \left( v \in V \right), a_{j} \right\}$ as in (2.1), 
where the integrals of stable abelian differentials around $a_{j}$ 
are defined as their residues at $x_{h}$ times $2 \pi \sqrt{-1}$ 
if $y_{h} = 0$ and $a_{j}$ is counter-clockwise oriented in the $\xi_{h}$-plane. 

{\rm (2)} 
For each $v \in V$, 
the restrictions $\omega_{v,i}|_{R_{0}}$ of $\omega_{v,i}$ to $R_{0}$ give rise to 
a basis of holomorphic abelian differentials on $R_{v}$ which is normalized for $\{ a_{v,i} \}_{i}$, 
and $\omega_{v,i}|_{R_{0}}$ become $0$ on $R_{v'}$ for $v' \in V - \{ v \}$. 
For each $j$, $\omega_{j}|_{R_{0}}$ has only poles which are simple at singular points 
on the closed path on $R_{0}$ corresponding to $b_{j}$. 
\end{it}
\vspace{2ex}

\noindent
{\it Proof.} 
By the above description of $\omega_{C/S}$, for each $v \in V$, 
there exists a unique basis $\{ (\omega_{v,i})_{0} \}$ of stable abelian differentials on $R_{0}$ 
which is normalized for $\left\{ a_{v,i} \left( v \in V \right), a_{j} \right\}$, 
namely it consists of the normalized basis of holomorphic abelian differentials on $R_{v}$ 
for $\left\{ a_{v,i} \right\}$, 
and $0$ on $R_{v'}$ $(v' \neq v)$. 
Similarly, for each $j$, 
there exists a unique stable abelian differential $(\omega_{j})_{0}$ on $R_{0}$ 
which is normalized for $\left\{ a_{v,i} \left( v \in V \right), a_{j} \right\}$. 
As is stated in \cite[Chapter 3, Section A]{HM}, 
the space of stable abelian differentials on $R_{0}$ has dimension $g$, 
and $\{ (\omega_{v,i})_{0} \left( v \in V \right), (\omega_{j})_{0} \}$ gives a basis of this space. 
Therefore, 
the coherent sheaf of stable abelian differentials on ${\mathcal R}_{C}/U_{r}$ 
is locally free on $U_{r}$ of rank $g$, 
and hence by the residue theorem, 
$\left\{ (\omega_{v,i})_{0}, (\omega_{j})_{0} \right\}$ can be uniquely extended to 
the normalized basis of stable abelian differentials on ${\mathcal R}_{C}/U_{r}$ as in (1). 
This implies the assertions (1) and (2). 
\ $\square$

\subsection{Variation of theta functions} 

Under the above notations, 
we consider the $1$-parameter family of complex curves 
$$
\{ R_{s} \}_{s \geq 0} \subset {\mathcal R}_{C}
$$ 
as a deformation of $R_{0}$ obtained by putting $y_{e} = s^{l(e)}$ $(e \in E)$, 
and take symplectic basis of $H_{1} \left( R_{s}, {\mathbb Z} \right)$ $(s > 0)$ as in (3.1) 
such that the underlying oriented paths in $R_{s}$ vary in $s$ continuously. 
Denote by $B_{C}$ the period matrix of the tropical curve $C$ 
which is defined in \cite{MZ} as a symmetric and positive definite bilinear form on 
$H_{1}(\Delta, {\mathbb Z}) \times H_{1}(\Delta, {\mathbb Z}) \cong 
{\mathbb Z}^{h_{1}(\Delta)} \times {\mathbb Z}^{h_{1}(\Delta)}$ 
given by $\langle b_{i}, b_{j} \rangle$, 
where 
$$
\langle h, h' \rangle = \left\{ \begin{array}{ll} 
l(h)   & (h = h'), \\ 
-l(h) & (h = -h'), \\ 
0     & (h \neq \pm h') 
\end{array} 
\right. 
$$   
for oriented edges $h, h' \in \pm E$. 
Then we describe the variation of the period matrices $B_{s}$ of $R_{s}$ for the basis (3.1) 
by $B_{C}$ and the period matrix $B_{v}$ of $R_{v}$ for the basis $\{ a_{v,i}, b_{v,i} \}_{i}$. 
Fix an identification 
$$
{\mathbb Z}^{g} \stackrel{\sim}{=} 
\left( \bigoplus_{v,i} {\mathbb Z} b_{v,i} \right) 
\bigoplus \left( \bigoplus_{j} {\mathbb Z} b_{j} \right) 
\stackrel{\sim}{=} 
\left( \bigoplus_{v \in V} {\mathbb Z}^{w(v)} \right) \bigoplus {\mathbb Z}^{h_{1}(\Delta)} 
\eqno(3.2) 
$$
which gives rise to the identifications 
$$
{\mathbb R}^{g} \stackrel{\sim}{=} 
\left( \bigoplus_{v,i} {\mathbb R} b_{v,i} \right) 
\bigoplus \left( \bigoplus_{j} {\mathbb R} b_{j} \right) 
\stackrel{\sim}{=} 
\left( \bigoplus_{v \in V} {\mathbb R}^{w(v)} \right) \bigoplus {\mathbb R}^{h_{1}(\Delta)} 
\eqno(3.3) 
$$
and 
$$
{\mathbb C}^{g} \stackrel{\sim}{=} 
\left( \bigoplus_{v,i} {\mathbb C} b_{v,i} \right) 
\bigoplus \left( \bigoplus_{j} {\mathbb C} b_{j} \right) 
\stackrel{\sim}{=} 
\left( \bigoplus_{v \in V} {\mathbb C}^{w(v)} \right) \bigoplus {\mathbb C}^{h_{1}(\Delta)}. 
\eqno(3.4) 
$$
Then we have: 
\vspace{2ex}

\noindent
{\bf Proposition 3.2.} 
\begin{it}

{\rm (1)} 
There exists the limit  
$\displaystyle \lim_{s \rightarrow +0} \frac{B_{s}}{\log s}$ 
which we denote by $\overline{B}$. 
Furthermore, 
$\overline{B}$ gives rise to a symmetric bilinear form on 
${\mathbb Z}^{g} \times {\mathbb Z}^{g}$ which becomes $0$ on
$$
\left( \bigoplus_{v \in V} {\mathbb Z}^{w(v)} \right) \times {\mathbb Z}^{h_{1}(\Delta)}, \ \ 
\left( \bigoplus_{v \in V} {\mathbb Z}^{w(v)} \right) \times 
\left( \bigoplus_{v \in V} {\mathbb Z}^{w(v)} \right), 
$$
and becomes $B_{C}/(2 \pi \sqrt{-1})$ on 
${\mathbb Z}^{h_{1}(\Delta)} \times {\mathbb Z}^{h_{1}(\Delta)} 
\stackrel{\sim}{=} H_{1}(\Delta, {\mathbb Z}) \times H_{1}(\Delta, {\mathbb Z})$. 

{\rm (2)} 
The limit  
$\displaystyle \lim_{s \rightarrow +0} \left( B_{s} - \log s \cdot \overline{B} \right)$ 
exists and gives rise to a symmetric bilinear form on 
${\mathbb Z}^{g} \times {\mathbb Z}^{g}$ which becomes 
\begin{itemize}

\item 
the period matrix $B_{v}$ of $\left( R_{v}, \left\{ a_{v, i}, b_{v, i} \right\} \right)$ on 
${\mathbb Z}^{w(v)} \times {\mathbb Z}^{w(v)}$, 

\item 
$0$ on ${\mathbb Z}^{w(v)} \times {\mathbb Z}^{w(v')}$ if $v \neq v'$, 

\item 
$\displaystyle \left( \int_{b_{j}|_{R_{v}}} \omega_{v,i}|_{R_{v}} \right)$ 
on ${\mathbb Z}^{w(v)} \times {\mathbb Z}^{h_{1}(\Delta)}$, 
where the integrals are taken along the restrictions of $b_{j}$ to $R_{v} \subset R_{0}$. 

\end{itemize}
Furthermore, 
the restriction $B_{0}$ of 
$\displaystyle \lim_{s \rightarrow +0} \left( B_{s} - \log s \cdot \overline{B} \right)$ 
to ${\mathbb Z}^{h_{1}(\Delta)} \times {\mathbb Z}^{h_{1}(\Delta)}$ 
is independent of the choice of the length function $l$ on $C$. 
\end{it} 
\vspace{2ex}

\noindent
{\it Proof.} 
By the description of $\omega_{v, i}|_{R_{0}}$ and $\omega_{j}|_{R_{0}}$ in Theorem 3.1 (2), 
the nonzero entries of $\displaystyle \lim_{s \rightarrow +0} \frac{2 \pi \sqrt{-1} B_{s}}{\log s}$ 
are finite sums of 
\begin{eqnarray*}
\pm \lim_{s \rightarrow +0} \frac{1}{\log s} \left( \int_{a}^{\xi_{h}} \frac{d \xi_{h}}{\xi_{h}} - 
\int_{\xi_{-h}}^{b} \frac{d \xi_{-h}}{\xi_{-h}} \right) 
& = & 
\pm \lim_{s \rightarrow +0} 
\frac{\log s \cdot l(|h|) - \log \left( \xi_{h}(a) \xi_{-h}(b) \right)}{\log s} 
\\
& = & 
\pm l(|h|),  
\end{eqnarray*}
where $h \in \pm E$ and $a \in R_{v_{h}}, b \in R_{v_{-h}}$ are close to $x_{h}, x_{-h}$ respectively. 
This implies (1). 
Furthermore, 
the nonzero entries of $B_{0}$ are ${\mathbb Z}$-linear sums of 
\begin{eqnarray*} 
\lefteqn{
\lim_{s \rightarrow +0} 
\left( \int_{a}^{\xi_{h}} \omega_{j}|_{R_{v_{h}}} - \int_{\xi_{-h}}^{b} \omega_{j}|_{R_{v_{-h}}} \right) 
- \log s \cdot l(|h|) 
} 
\\ 
& = & 
\int_{a}^{x_{h}} f(\xi_{h}) d \xi_{h} - \int_{x_{-h}}^{b} g(\xi_{-h}) d \xi_{-h} - 
\frac{\log \left( \xi_{h}(a) \xi_{-h}(b) \right)}{2 \pi \sqrt{-1}} 
\end{eqnarray*}
for points $a \in R_{v_{h}}, b \in R_{v_{-h}}$ if the differentials $\omega_{j}$ 
on ${\mathcal R}_{C}/U_{r}$ given in Theorem 3.1 satisfy 
$$
\omega_{j}|_{R_{0}} = \left\{ \begin{array}{ll} 
{\displaystyle \left( \frac{1}{2 \pi \sqrt{-1} \, \xi_{h}} + f(\xi_{h}) \right) d \xi_{h}} 
& \mbox{at $x_{h}$,} 
\\ 
{\displaystyle \left( \frac{1}{2 \pi \sqrt{-1} \, \xi_{-h}} + g(\xi_{-h}) \right) d \xi_{-h}} 
& \mbox{at $x_{-h}$,} 
\end{array} \right. 
$$ 
where $f(\xi_{h}), g(\xi_{-h})$ are holomorphic at $x_{h}, x_{-h}$ respectively. 
Therefore, $B_{0}$ is independent of $l$, 
and the remaining assertions in (2) follow from Theorem 3.1. 
\ $\square$ 
\vspace{2ex}

Since $B_{C}$ is positive definite, 
for an element $\alpha$ of $H_{1}(\Delta, {\mathbb R})$, 
one can define the associated tropical theta function 
$$
\Theta_{B_{C}}(\alpha) = 
\max \left\{ \left. \alpha B_{C} x^{T} - \frac{1}{2} x B_{C} x^{T} \, \right| 
x \in H_{1}(\Delta, {\mathbb Z}) \right\}, 
$$
and the Delaunay set 
$$
D_{\alpha, B_{C}} = 
\left\{ x \in H_{1}(\Delta, {\mathbb Z}) \, \left| \ 
\alpha B_{C} x^{T} - \frac{1}{2} x B_{C} x^{T} = \Theta_{B_{C}}(\alpha) \right. \right\} 
$$
which is a finite set (cf. \cite{AgCSS, AgFMS, FoRSS, MZ}). 
Then the following theorem was shown as \cite[Theorem 5.2]{I2} 
when $C$ has trivial weights. 
\vspace{2ex}

\noindent
{\bf Theorem 3.3.} 
\begin{it} 
Let $\Theta_{(R_{v}, \{ a_{v,i}, b_{v,i} \})}$ denote the Riemann theta function defined for 
$\left( R_{v}, \{ a_{v,i}, b_{v,i} \} \right)$ given in (2.2) if $w(v) > 0$, 
and the constant $1$ if $w(v) = 0$. 
For $\overline{z} = \left( (z_{v})_{v}, z \right) \in {\mathbb C}^{g}$ and 
$\overline{\alpha} = \left( (\alpha_{v})_{v}, \alpha \right)\in {\mathbb R}^{g}$ 
under (3.4) and (3.3) respectively, 
the regularized tropical limit of $\Theta_{(R_{s}, \{ a_{v,i}, a_{j}, b_{v,i}, b_{j} \})}$ defined as 
$$
\lim_{s \rightarrow +0} s^{\Theta_{B_{C}}(\alpha)} \cdot 
\Theta_{(R_{s}, \{ a_{v,i}, a_{j}, b_{v,i}, b_{j} \})} 
\left( \overline{z} - \log s \cdot \overline{\alpha} \overline{B} \right) 
\eqno(3.5) 
$$ 
exists and becomes  
$$
\sum_{x \in D_{\alpha, B_{C}}} 
\exp \left( \pi \sqrt{-1} \left( x B_{0} x^{T} + 2 z x^{T} \right) \right) 
\prod_{v \in V} \Theta_{(R_{v}, \{ a_{v,i}, b_{v,i} \})} 
\left( z_{v} + \int_{x|_{R_{v}}} \omega_{v,i}|_{R_{v}} \right), 
$$
where 
$\int_{x|_{R_{v}}} \omega_{v,i}|_{R_{v}} = \sum_{j} x_{j} \int_{b_{j}|_{R_{v}}} \omega_{v,i}|_{R_{v}}$ 
if $x$ is a ${\mathbb Z}$-linear sum $\sum_{j} x_{j} b_{j}$ of $b_{j}$ 
in $H_{1}(\Delta, {\mathbb Z})$. 
\end{it} 
\vspace{2ex}

\noindent
{\it Proof.} 
Since $- \Theta_{B_{C}}(\alpha)$ is the lowest exponent of $s$ 
in all the terms of the absolutely convergent series 
\begin{eqnarray*}
\lefteqn{
\Theta_{(R_{s}, \{ a_{v,i}, a_{j}, b_{v,i}, b_{j} \})} 
\left( \overline{z} - \log s \cdot \overline{\alpha} \overline{B}  \right)
} 
\\
& = & 
\sum_{\overline{u} \in {\mathbb Z}^{g}} 
\exp \left( \pi \sqrt{-1} \left( \overline{u} B_{s} \overline{u}^{T} + 
2 \left( \overline{z} - \log s \cdot \overline{\alpha} \overline{B} \right) 
\overline{u}^{T} \right) \right),
\end{eqnarray*} 
by identifying $\overline{u} \in {\mathbb Z}^{g}$ with $\left( (u_{v})_{v}, u \right)$ 
under (3.2), 
Proposition 3.2 implies that (3.5) becomes the finite sum of 
\begin{eqnarray*}
\lefteqn{
\exp \left( \pi \sqrt{-1} \left( u B_{0} u^{T} + 2 z u^{T} \right) \right) 
}
\\ 
& & 
\cdot 
\prod_{v \in V} \sum_{u_{v}} 
\exp \left( \pi \sqrt{-1} \left( (u_{v})_{v} B_{v} (u_{v})_{v}^{T} + 
2 \left( z_{v} + \int_{u|_{R_{v}}} \omega_{v,i}|_{R_{v}} \right) (u_{v})_{v}^{T} \right) \right), 
\end{eqnarray*}
where $u$ runs through the finite Delaunay set $D_{\alpha, B_{C}}$. 
This implies the assertion. 
\ $\square$

\subsection{Variation of KP solutions} 

Fix a vertex $v_{0} \in V$, a point $p$ on the irreducible component $R_{v_{0}}$ of $R_{0}$ 
and a local coordinate $z$ at $p$ such that $z(p) = 0$ 
which give a point and a local coordinate respectively of each member of ${\mathcal R}_{C}$. 
\vspace{2ex}  

\noindent
{\bf Theorem 3.4.} 
\begin{it}

{\rm (1)} 
There exists a unique meromorphic section $\omega^{(n)}$ 
of $\omega_{{\mathcal R}_{C}/U_{r}}$ such that  
\begin{itemize}

\item 
$\omega^{(n)}$ is holomorphic outside $p$, 

\item 
$\displaystyle \omega^{(n)} - \frac{dz}{z^{n+1}}$ is holomorphic at $p$, 

\item 
$\displaystyle \int_{a_{v,i}} \omega^{(n)} = \displaystyle \int_{a_{j}} \omega^{(n)} = 0$ 
for all $a_{v,i}$ and $a_{j}$, 
where the integrals are defined as in Theorem 3.1 (1) for 
singular members in ${\mathcal R}_{C}$. 

\end{itemize}

{\rm (2)} 
The restriction $\omega^{(n)}|_{R_{0}}$ of $\omega^{(n)}$ to $R_{0}$ becomes 
the unique meromorphic abelian differential on $R_{v_{0}}$ 
with unique pole at $p$ of order $n + 1$ 
which is normalized for $\left( \{ a_{v_{0}, i} \}, p, z \right)$, 
and it becomes $0$ on $R_{v}$ for $v \in V - \{ v_{0} \}$. 
\end{it}
\vspace{2ex}

\noindent
{\it Proof.} 
By the description of $\omega_{C/S}$, 
there exists a meromorphic abelian differential $(\omega^{(n)})_{0}$ on $R_{0}$ 
such that $(\omega^{(n)})_{0}|_{R_{v_{0}}}$ has a unique pole at $p$ of order $n + 1$ 
and is normalized for $\left( \{ a_{v_{0}, i} \}, p, z \right)$, 
and that $(\omega^{(n)})_{0}|_{R_{v}} = 0$ for $v \neq v_{0}$. 
Furthermore, by the Riemann-Roch theorem for nodal curves stated in 
\cite[Chapter 3, Section A]{HM}, 
the dimension of the space consisting of meromorphic sections of $\omega_{R_{0}}$ 
with unique pole at $p$ of order $\leq n + 1$ becomes 
\begin{eqnarray*}
& & 
\deg \left( \omega_{R_{0}}((n+1) p) \right) - g + 1 + 
\dim H^{0} \left( R_{0}, {\mathcal O}_{R_{0}}(-(n+1) p) \right) 
\\
& = & 
(2g - 2 + n + 1) - g + 1 
\\
& = & 
g + n. 
\end{eqnarray*}
Therefore, the coherent sheaf of meromorphic sections of $\omega_{{\mathcal R}_{C}/U_{r}}$ 
with unique pole at $p$ of order $\leq n+1$ is locally free on $U_{r}$ of rank $g + n$, 
and hence $(\omega^{(n)})_{0}$ can be uniquely extended to 
the meromorphic section of $\omega_{{\mathcal R}_{C}/U_{r}}$ as in (1). 
This implies the assertions (1) and (2). 
\ $\square$ 
\vspace{2ex}

Let ${\mathcal X} = (R_{s}, \{ a_{v,i}, a_{j}, b_{v,i}, b_{j} \}, p, z)$ denote 
the family of dressed Riemann surfaces, 
and consider the regularized tropical limit of the tau functions given in (2.3) 
associated with ${\mathcal X}$. 
Then the following theorem was shown as \cite[Theorem 5.3]{I2} 
when $C$ has trivial weights. 
\vspace{2ex}

\noindent 
{\bf Theorem 3.5.}
\begin{it}
For the $\omega_{j}$ and $\omega^{(n)}$ given in Theorems 3.1 and 3.4 respectively, 
take $r_{m} \in {\mathbb C}^{h_{1}(\Delta)}$ and $q_{n, m} \in {\mathbb C}$ 
which are defined as 
$$
\left( \omega_{j}|_{R_{0}} \right)_{j} = \sum_{m=1}^{\infty} r_{m} z^{m-1} dz \ 
\mbox{at $p$}, \ \  
\left( \omega^{(n)}|_{R_{0}} \right) = 
\left( \frac{1}{z^{n+1}} + \sum_{m=1}^{\infty} \frac{q_{n, m}}{n} z^{m-1} \right) dz \ 
\mbox{at $p$.} 
$$  
For $\overline{c} = \left( (c_{v})_{v}, c \right) \in {\mathbb C}^{g}$ under (3.4), 
put $\overline{c}' = \overline{c} - \log s \cdot \overline{\alpha} B_{s}$. 
Then the limit under $s \rightarrow +0$ of 
$s^{\Theta_{B_{C}}(\alpha)} \cdot \tau \left(t, {\mathcal X}_{\overline{c}'} \right)$ becomes  
\begin{eqnarray*}
\lefteqn{
\sum_{x \in D_{\alpha, B_{C}}} 
\exp \left( \pi \sqrt{-1} \left( x B_{0} x^{T} + 
2 \left( c + \sum_{m=1}^{\infty} r_{m} t_{m} \right) x^{T}  \right) \right) 
}
\\
& & \cdot \ 
\tau \left( t, \left( X_{v_{0}} \right)_{c'_{v_{0}}} \right) 
\prod_{v \in V - \{ v_{0} \}} 
\Theta_{(R_{v}, \{ a_{v,i}, b_{v,i} \})} \left( c'_{v} \right), 
\end{eqnarray*}
where $c'_{v} = c_{v} + \int_{x|_{R_{v}}} \omega_{v,i}|_{R_{v}}$ and 
$X_{v_{0}} = \left( R_{v_{0}}, \left\{ a_{v_{0}, i}, b_{v_{0}, i} \right\}, p, z \right)$. 
Furthermore, 
this function of $t$ gives a solution to the KP hierarchy given in (2.4)--(2.6). 
\end{it}
\vspace{2ex}

\noindent
{\it Proof.} 
The assertion follows from Theorems 3.3, 3.4 (2) and that 
$s^{\Theta_{B_{C}}(\alpha)} \cdot \tau \left(t, {\mathcal X}_{\overline{c}'} \right)$, 
$\tau \left(t, {\mathcal X}_{\overline{c}'} \right)$ 
give the same micro-differential operators in (2.5) and (2.6). 
\ $\square$

\subsection{Decomposition of KP solutions}  

We consider decompositions of KP solutions given in Theorem 3.5 to their components.  
Let $\Delta = (V, E)$ be a connected graph, 
and $t_{e} = t_{-e}$ $(e \in E)$ be variables. 
Then the period matrix $B_{\Delta} = (B_{ij})_{1 \leq i, j \leq h_{1}(\Delta)}$ of $\Delta$ 
is defined as the symmetric bilinear form on 
$H_{1}(\Delta, {\mathbb Z}) \times H_{1}(\Delta, {\mathbb Z})$ 
with values in $\bigoplus_{e \in E} {\mathbb Z} \cdot t_{e}$ satisfying 
$$
\langle h, h' \rangle = \left\{ \begin{array}{ll} 
t_{h}    & (h = h'), \\ 
- t_{h} & (h = -h'), \\ 
0        & (h \neq \pm h') 
\end{array} 
\right. 
$$   
for oriented edges $h, h' \in \pm E$. 
Since $B_{C} = B_{\Delta}|_{t_{e} = l(e)}$ is positive definite 
for any tropical curve $C$ associated with $\Delta$ and a length function $l$, 
if $x$ is a nonzero element of $H_{1}(\Delta, {\mathbb Z})$, 
then $x B_{\Delta} x^{T}$ is a ${\mathbb Z}$-linear sum of $t_{e}$ $(e \in E)$ 
whose coefficients are all positive. 
For an element $\alpha$ of $H_{1}(\Delta, {\mathbb R})$, 
we call an element of 
$\bigoplus_{e \in E} {\mathbb R} \cdot t_{e}$ {\it maximal} for $\alpha$ 
if this is a maximal element of 
$$
\left\{ \left. \alpha B_{\Delta} x^{T} - \frac{1}{2} x B_{\Delta} x^{T} \ \right| 
x \in H_{1}(\Delta, {\mathbb Z}) \right\} 
\subset \bigoplus_{e \in E} {\mathbb R} \cdot t_{e}
$$ 
with respect to the partial order defined as 
$$
\sum_{e \in E} a_{e} t_{e} \leq \sum_{e \in E} b_{e} t_{e} \ \Leftrightarrow \ 
a_{e} \leq b_{e} \ \mbox{for all $e \in E$.} 
$$
For a maximal element 
$a = \sum_{e \in E} a_{e} t_{e} \in \bigoplus_{e \in E} {\mathbb R} \cdot t_{e}$ 
for $\alpha$, 
the associated Delaunay set is defied as 
$$
D_{\alpha, a} = \left\{ x \in H_{1}(\Delta, {\mathbb Z}) \left| \ 
\alpha B_{\Delta} x^{T} - \frac{1}{2} x B_{\Delta} x^{T} = a \right. \right\}
$$
which is a finite set, 
and for the nodal complex curve $R_{0} = \bigcup_{v \in V} R_{v}$, 
$\tau_{\alpha, a}(t)$ is defined as 
\begin{eqnarray*}
\lefteqn{
\sum_{x \in D_{\alpha, a}} 
\exp \left( \pi \sqrt{-1} \left( x B_{0} x^{T} + 
2 \left( c + \sum_{m=1}^{\infty} r_{m} t_{m} \right) x^{T}  \right) \right) 
}
\\
& & \cdot \ 
\tau \left( t, \left( X_{v_{0}} \right)_{c'_{v_{0}}} \right) 
\prod_{v \in V - \{ v_{0} \}} 
\Theta_{(R_{v}, \{ a_{v,i}, b_{v,i} \})} \left( c'_{v} \right). 
\end{eqnarray*}

\noindent
{\bf Theorem 3.6.}
\begin{it}

{\rm (1)} 
Let $\alpha$ be an element of $H_{1}(\Delta, {\mathbb R})$. 
Then for a tropical curve $C$ associated with $\Delta$ and a length function $l$, 
$\Theta_{B_{C}}(\alpha)$ is equal to the maximum value of $\sum_{e \in E} a_{e} l(e)$, 
where $a = \sum_{e \in E} a_{e} t_{e}$ runs through maximal elements of 
$\bigoplus_{e \in E} {\mathbb R} \cdot t_{e}$ for $\alpha$. 
Furthermore, the finite sum of $\tau_{\alpha, a}(t)$ for maximal elements 
$a = \sum_{e \in E} a_{e} t_{e}$ for $\alpha$ satisfying 
$$
\sum_{e \in E} a_{e} l(e) = \Theta_{B_{C}}(\alpha) 
$$
gives a solution to the KP hierarchy. 

{\rm (2)} 
Any solution to the KP hierarchy given in Theorem 3.5 can be obtained as in (1). 
\end{it}  
\vspace{2ex}

\noindent
{\it Proof.} 
The assertion follows from Theorem 3.5. 
\ $\square$ 
\vspace{2ex}

\noindent
{\bf Acknowledgments}  
The author would like to thank the anonymous referee for helpful remarks and suggestions 
in modifying this manuscript. 
This work is partially supported by the JSPS Grant-in-Aid for 
Scientific Research No. 20K03516.

\end{document}